\documentclass[twocolumn,showpacs,preprintnumbers, nofootinbib,superscriptaddress, amsmath,amssymb]{revtex4}
\usepackage{amsthm}
\usepackage{bm}
\usepackage{epsfig}

\DeclareFontFamily{OT1}{rsfs}{}
\DeclareFontShape{OT1}{rsfs}{m}{n}{<5> rsfs5 <7> rsfs7 <10> rsfs10}{}
\DeclareSymbolFont{mathrsfs}{OT1}{rsfs}{m}{n}
\DeclareSymbolFontAlphabet{\mathrsfs}{mathrsfs}

\newcommand*{\ket}[1]{\left| #1 \right\rangle}
\newcommand*{\bra}[1]{\left\langle{#1}\right|}

\newcommand{\intsum}{ \mathop{\hbox to4pt{ $\sum$ \hss}{\displaystyle\int}} }

\begin{document}

\title{Enhancement and suppression of tunneling by controlling symmetries of a potential barrier} 

\author{Denys I. Bondar}
\email{dbondar@sciborg.uwaterloo.ca}
\affiliation{University of Waterloo, Waterloo, Ontario N2L 3G1, Canada}
\affiliation{National Research Council of Canada, Ottawa, Ontario K1A 0R6, Canada}

\author{Wing-Ki Liu}
\email{wkliu@sciborg.uwaterloo.ca}
\affiliation{University of Waterloo, Waterloo, Ontario N2L 3G1, Canada}
\affiliation{Department of Physics, The Chinese University of Hong Kong,
Shatin, NT, Hong Kong}

\author{Misha Yu. Ivanov}
\email{m.ivanov@imperial.ac.uk}
\affiliation{Imperial College, London SW7 2BW, U.K.}

\begin{abstract}
We present a class of 2D systems which shows a counterintuitive property that contradicts a semi classical intuition: A 2D quantum particle ``prefers'' tunneling through a barrier rather than traveling above it. Viewing the one particle 2D system as the system of two 1D particles, it is demonstrated that this effect occurs due to a specific symmetry of the barrier that forces excitations of the interparticle degree of freedom that, in turn, leads to the appearance of an effective potential barrier even though there is no ``real'' barrier. This phenomenon cannot exist in 1D.
\end{abstract}

\pacs{03.65.Xp, 32.80.Qk}

\maketitle

\section{Introduction}

Quantum tunneling has been one of the most important problems in quantum mechanics since its foundation. The simplest problems of tunneling are one-dimensional, which is where our intuition on tunneling comes from. The extension of 1D tunneling to many dimensions is not straightforward. There are many peculiarities that appear in many dimensional cases that do not exist in 1D (for systematic studies of such differences see, e.g., Refs. \cite{Chabanov1999, Chabanov2000, Zakhariev2008}). Quite often many dimensional tunneling is equated to the tunneling of complex (i.e., many particle) systems.  

Key aspects of the quantum mechanical tunneling of complex systems were analyzed by Zakhariev {\it et al.} \cite{Zakhariev1964, Amirkhanov1966} in the mid-1960s; nevertheless, this problem has become an area of active research only in the past few decades (see, e.g., Refs. \cite{Tomsovic1998, Takagi2002, Zakhariev2002, Razavy2003a, Ankerhold2007} and references therein). Tunneling of a diatomic molecule has been studied in Refs. \cite{Goodvin2005, Goodvin2005a, Lee2006, Hnybida2008, Shegelski2008, Kavka2010}. Mechanisms of single and double proton transfer have been modelled by multidimensional tunneling \cite{Smedarchina1995, Smedarchina2007, Smedarchina2008}.  Time-dependent numerical study of tunneling dynamics of a two-particle quantum system with an internal degree of freedom has been analyzed in Ref. \cite{Volkova2006}, and an enhancement of the tunneling probability due to the formation of a long-lived resonant state of the system in the barrier region has been discovered (similar analytical studies have been done in Ref. \cite{Yamamoto1996}). It has also been suggested that collective tunneling of electrons may have an important contribution to multiple ionization of atoms in a superstrong laser field \cite{Kornev2009}. Quantum tunneling of complex systems is not only of theoretical interest. Recent experiments where this phenomenon is observed directly include tunneling of a singe hydrogen atom \cite{Lauhon2000}, resonant tunneling of Cooper pairs \cite{Toppari2007}, and a bosonic Josephson junction consisting of two weakly coupled Bose-Einstein condensates in a macroscopic double-well potential \cite{Albiez2005}.

We present a class of 2D systems which has a counterintuitive property that contradicts the semi classical intuition: A 2D quantum particle ``prefers'' tunneling through to flying above a barrier. According to our analysis, such ``paradoxical'' dynamics is caused by a peculiar symmetry of the barrier that leads to excitations of an interparticle degree of freedom. There is no 1D counterpart of such systems.  

The rest of the article is organized as follows: In Sec. \ref{Sec2}, we present the systems and describe the counterintuitive effect. The observed ``paradox'' is explained in Sec. \ref{Sec3}. Connections between the phenomenon and classical physics are discussed in Sec. \ref{Sec4}. Concluding remarks and a possible application of the effect to quantum control are presented in the last section.

\section{Formulation of the ``paradox''}\label{Sec2}

\begin{figure}
\begin{center}
\includegraphics[scale=0.6]{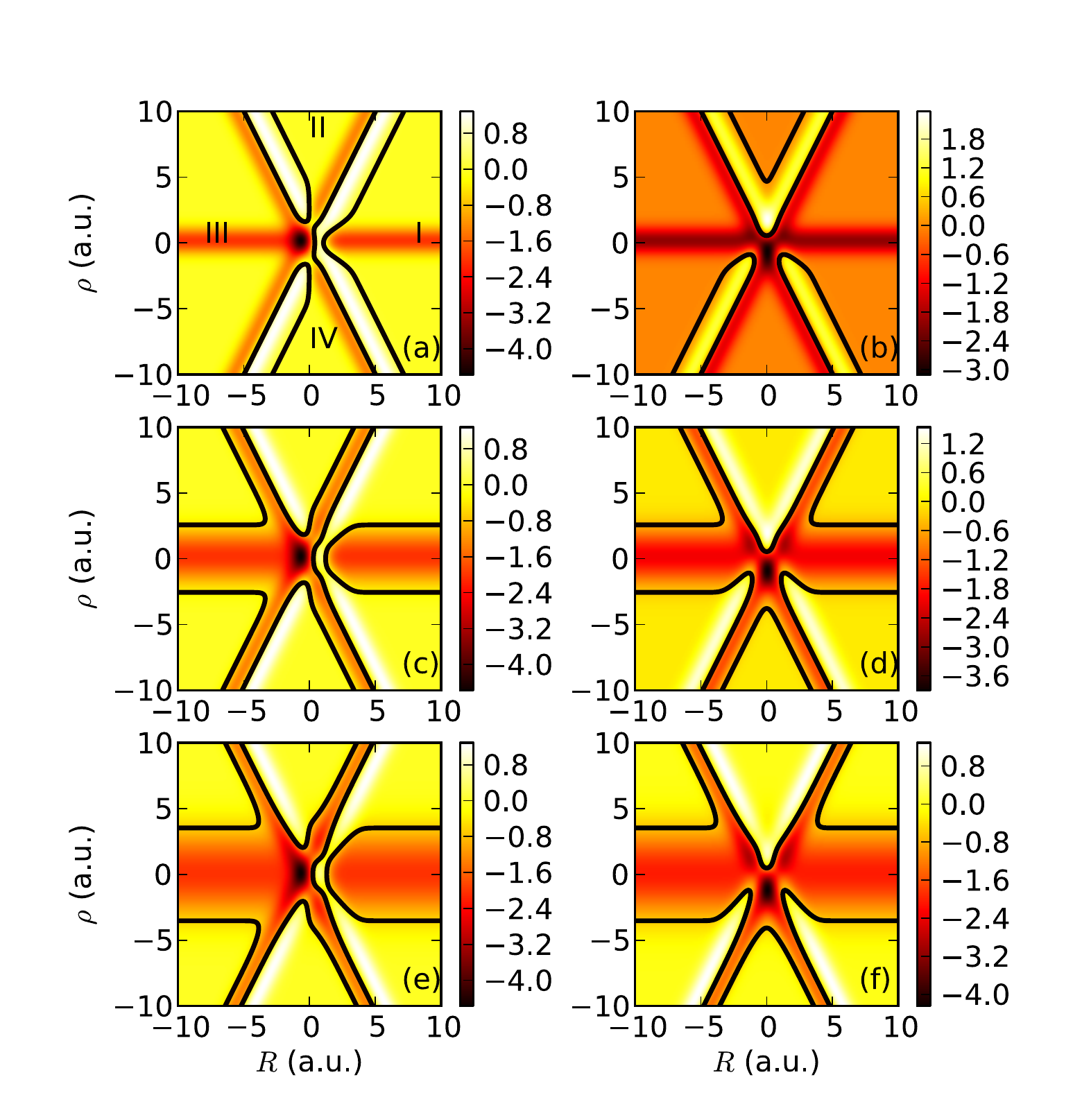}
\caption{(Color online) Plots of the potential  $\Omega_N (\alpha; R, \rho)$ [Eq. (\ref{PotentialDeff})] for different $\alpha$ and $N$. Roman numerals in plot (a) label quadrants. Black solid lines denote the level sets of the expectation value of the initial energy, $\bar{E}_N$ [Eq. (\ref{AvEDef})], i.e., the boundary between the classically allowed and classically forbidden regions. (a) $N=1$ and $\alpha=3$; (b) $N=1$ and $\alpha = -3$; (c) $N=2$ and $\alpha=3$; (d) $N=2$ and $\alpha=-3$; (e) $N=4$ and $\alpha=3$; (f) $N=4$ and $\alpha=-3$. }\label{Fig_potentials}
\end{center}
\end{figure}

Let us consider a particle moving in 2D (coordinates $x_1$ and $x_2$) toward a barrier located at the origin $x_1=x_2=0$. The initial velocity of the particle is chosen to be directed along the diagonal $x_1=x_2$, incident on the barrier from the third (where $x_1 < 0$ and $x_2 < 0$) to the first quadrant (where $x_1 > 0$ and $x_2 > 0$); see Fig. \ref{Fig_potentials}. While the numerical calculations are done for a specific Hamiltonian, the analytical analysis that follows relies exclusively on the symmetry properties of the 2D potential, making our conclusions, drawn from the numerical analysis, general.

The model Hamiltonian for our system is chosen as (atomic units are used throughout)
\begin{eqnarray}
&& \hat{H}_N (\alpha) = -\frac 1{2} \left( \frac{\partial^2}{\partial x_1^2} +   \frac{\partial^2}{\partial x_2^2} \right) + \Omega_N (\alpha; x_1, x_2), \label{HamiltonianCartesianDef} \\
&& \Omega_N (\alpha; x_1, x_2) = \alpha V(x_1) + 3V(x_2) + U_N(x_2 - x_1). 
\end{eqnarray}
where $N=1,2,4$ and $\alpha$ being an arbitrary real parameter. The potentials $V(x_1)$ and $V(x_2)$ describe the potential barriers near the origin, for the motion along each of the two coordinates. The parameter $\alpha$ allows us to vary the relative height of the barriers. We have chosen 
\begin{eqnarray}
V(x) = x \exp(-x^2),
\end{eqnarray}
which corresponds to a potential barrier preceded by a potential well.

The potential $U_N(\rho)$ describes the coupling between the two degrees of freedom. In the absence of $U_N(\rho)$ [i.e., for $U_N(\rho)=0$], the 2D dynamics breaks into two uncoupled 1D motions. Nontrivial features in tunneling appear as the result of nonzero coupling of the two degrees of freedom. 

Before we describe the choice of $U_N(\rho)$ in our model, let us introduce  the center of mass ($R$) and relative ($\rho$) coordinates
\begin{eqnarray}\label{SimpleCMandRelativeCoord}
R = (x_1 + x_2)/2, \qquad \rho = x_2 -x_1.
\end{eqnarray}
 The Hamiltonian (\ref{HamiltonianCartesianDef}) in these new coordinates reads 
 \begin{eqnarray}\label{Hamiltonian_NewCoord}
\hat{H}_N(\alpha) &=& \frac{-1}{2M}\frac{\partial^2}{\partial R^2} + \frac{-1}{2\mu}\frac{\partial^2}{\partial \rho^2} +  \Omega_N (\alpha; \rho, R), \nonumber\\
\Omega_N (\alpha; R, \rho) &=& \alpha V\left( R - \rho/2\right) + 3V\left( R +  \rho/2\right) \nonumber\\
&&+ U_N(\rho), \label{PotentialDeff}
\end{eqnarray}
where $\mu = 1/2$ and $M = 2$. 

Now we can specify the potential that couples the two degrees of freedom and see its role in the problem. If $U_N(\rho)$ is attractive, as it is in our calculations, it may support bound states. These bound states, and their symmetries, play a key role.

Here, we set $U_N$ to describe a short-range attraction,
\begin{eqnarray}\label{Un_potentials}
U_N (\rho) = -A\exp\left( -\rho^2/r_N^2 \right).
\end{eqnarray}
Varying the parameter $r_N$, we change the number of bound states supported by the attracting potential. In the calculations, we use $A=2$ and $r_1 = 1$, $r_2 = 1.961$,  $r_4 =3.162$ corresponding to one, two, and four bound states supported by the Hamiltonian $-1/(2\mu) \partial^2/\partial \rho^2 + U_N(\rho)$. The energies of these states are $-0.955$ for $U_1$; $-1.377$ and $-0.372$ for $U_2$; and, finally, $-1.590$, $-0.856$, $-0.308$, and $-0.012$ for $U_4$.

Following Ref. \cite{Volkova2006}, we study the tunneling within the time-dependent approach solving the time-dependent Schr\"{o}dinger equation,
\begin{eqnarray}\label{TimDepSchEq}
\left[ i\partial/\partial t - \hat{H}_N(\alpha)\right] \Psi_N(\alpha; t, x_1, x_2) = 0,
\end{eqnarray}
with the initial condition at $t=0$ that reads in the coordinates (\ref{SimpleCMandRelativeCoord}) as
\begin{eqnarray}\label{InitialCondition}
\Psi_N(\alpha; 0, R, \rho) = C \phi_g(\rho) e^{ -(R-\bar{R})^2/\left(2\sigma_R^2\right) + i\sqrt{2M E_{cm}}R  }.
\end{eqnarray}
Here $C$ is a normalization constant and $\phi_g(\rho)$ is the ground state of the interparticle Hamiltonian, $-1/(2\mu) \partial^2/\partial \rho^2 + U_N(\rho)$. In all our studies, we set $m=1$, $\bar{R} = -55$, $\sigma_R = 3$, and $E_{cm} = 1$ (all values are in atomic units). 

The initial wave function (\ref{InitialCondition}) is localized in the third quadrant, and we calculate the probability of finding the particle in the first quadrant, i.e., the probability of tunneling at later time $\tau$. The reason for using $\phi_g(\rho)$ as the relative coordinate part of the initial wave function is that we wanted to avoid spreading of the wave packet along $\rho$ before it reached the potential barrier.

We also present the initial expectation value of energy
\begin{eqnarray}\label{AvEDef}
\bar{E}_N = \bra{\Psi_N(\alpha; 0, x_1, x_2) }\hat{H}_N(\alpha) \ket{\Psi_N(\alpha; 0, x_1, x_2)}, \\
\bar{E}_1 = 0.05911, \quad \bar{E}_2 = -0.3631, \quad \bar{E}_4 = -0.5766 \nonumber
\end{eqnarray}
(all values are in atomic units). Rigorously speaking,  $\bar{E}_N$ depends on $\alpha$; however, this dependence is very weak because the initial wave function (\ref{InitialCondition}), independent of $\alpha$, is mostly localized in the region where the potential barrier, $\alpha V(x_1) + 3V(x_2)$,  vanishes.

The probabilities of tunneling, disintegration (see below for the clarification of this term), and reflection are defined as follows: 
\begin{eqnarray}
P_T^{(N)} (\alpha, \tau) &=& \int_0^{\infty} dx_1 \int_0^{\infty} dx_2 \, \left| \Psi_N(\alpha; \tau, x_1, x_2) \right|^2,  \label{Prob_T_deff}\\
P_D^{(N)} (\alpha, \tau) &=& \int_{-\infty}^0 dx_1 \int_0^{\infty} dx_2 \, \left| \Psi_N(\alpha; \tau, x_1, x_2) \right|^2 \nonumber\\
	&+& \int_0^{\infty} dx_1 \int_{-\infty}^0 dx_2 \, \left| \Psi_N(\alpha; \tau, x_1, x_2) \right|^2, \label{Prob_D_deff}\\
P_R^{(N)} (\alpha, \tau) &=& \int_{-\infty}^0 dx_1 \int_{-\infty}^0 dx_2 \, \left| \Psi_N(\alpha; \tau, x_1, x_2) \right|^2. \label{Prob_R_deff}
\end{eqnarray}

However, since the potential barrier, $\alpha V(x_1) + 3V(x_2)$, has ``well''  and ``hill'' regions, we also employ the corresponding ``shifted'' probabilities to exclude regions were the potential barrier is localized
\begin{eqnarray}
p_t^{(N)} (\alpha, \tau) &=& \int_3^{\infty} dx_1 \int_3^{\infty} dx_2 \, \left| \Psi_N(\alpha; \tau, x_1, x_2) \right|^2, \label{prob_t_deff} \\
p_d^{(N)} (\alpha, \tau) &=& \int_{-\infty}^{-3} dx_1 \int_3^{\infty} dx_2 \, \left| \Psi_N(\alpha; \tau, x_1, x_2) \right|^2 \nonumber\\
	&+& \int_3^{\infty} dx_1 \int_{-\infty}^{-3} dx_2 \, \left| \Psi_N(\alpha; \tau, x_1, x_2) \right|^2, \label{prob_d_deff}\\
p_r^{(N)} (\alpha, \tau) &=& \int_{-\infty}^{-3} dx_1 \int_{-\infty}^{-3} dx_2 \, \left| \Psi_N(\alpha; \tau, x_1, x_2) \right|^2, \label{prob_r_deff}\\
p_s^{(N)} (\alpha, \tau) &=& 1 - p_t^{(N)}  - p_d^{(N)}  - p_r^{(N)},
\end{eqnarray}
$p_s^{(N)}$ is the probability that a particle is trapped in the neighborhood of the potential barrier (in fact, mostly in the well region of the potential barrier). We introduce these quantities to verify that our conclusions are not due to variations of the probability density in a neighborhood  of the potential barrier (see Ref. \cite{EPAPS_Animations}).

The Hamiltonian (\ref{HamiltonianCartesianDef}) can be viewed as the Hamiltonian of two 1D particles, where $x_{1,2}$ are coordinates of the first and second particles, respectively.  This interpretation is crucial to explain the observed effect. Utilizing such a point of view, quantities $P_D^{(N)}$ [Eq. (\ref{Prob_D_deff})] and $p_d^{(N)}$ [Eq. (\ref{prob_d_deff})] can be indeed labeled as the probabilities of disintegration because the particles are flying apart (i.e., the two particle system is disintegrating) if after sufficiently long time $\tau$ either $x_1>0$ and $x_2 <0$ or $x_1<0$ and $x_2>0$.

Before stating the results of numerical calculations, let us qualitatively analyze dynamics of the system within a semiclassical consideration. Figure \ref{Fig_potentials} presents the plots of the potentials (\ref{PotentialDeff}). According to the initial condition [see Eqs. (\ref{InitialCondition}) and (\ref{AvEDef})], the particle is located on the axis $\rho=0$ and its initial velocity is directed along this axis toward the first quadrant, and the amplitude of the velocity is chosen such that the total energy of the particle equals $\bar{E}_N$; hence, the boundaries between the classically allowed and classically forbidden regions are drawn by solid black lines in Fig. \ref{Fig_potentials}. Now compare Fig. \ref{Fig_potentials}(a) with Fig. \ref{Fig_potentials}(b). Since the semi classical counterpart of our quantum particle ``experiences'' the barrier in Fig. \ref{Fig_potentials}(a) (penetration though a barrier is of exponentially small probability) and does not ``feel'' any barrier in Fig. \ref{Fig_potentials}(b) (the particle moves solely in the classically allowed region) while traveling along the axis $\rho=0$, then one would intuitively conclude that the probability of finding the particle in the first quadrant in Fig. \ref{Fig_potentials}(a) ought to be smaller than in Fig. \ref{Fig_potentials}(b). By the same token, the very same probabilities in Figs. \ref{Fig_potentials}(c) and \ref{Fig_potentials}(e) should be smaller than in Figs. \ref{Fig_potentials}(d) and \ref{Fig_potentials}(f), respectively. Further discussions of the phenomenon from the point of view of classical trajectories are presented in Sec. \ref{Sec4}.

\begin{figure}
\begin{center}
\includegraphics[scale=0.5]{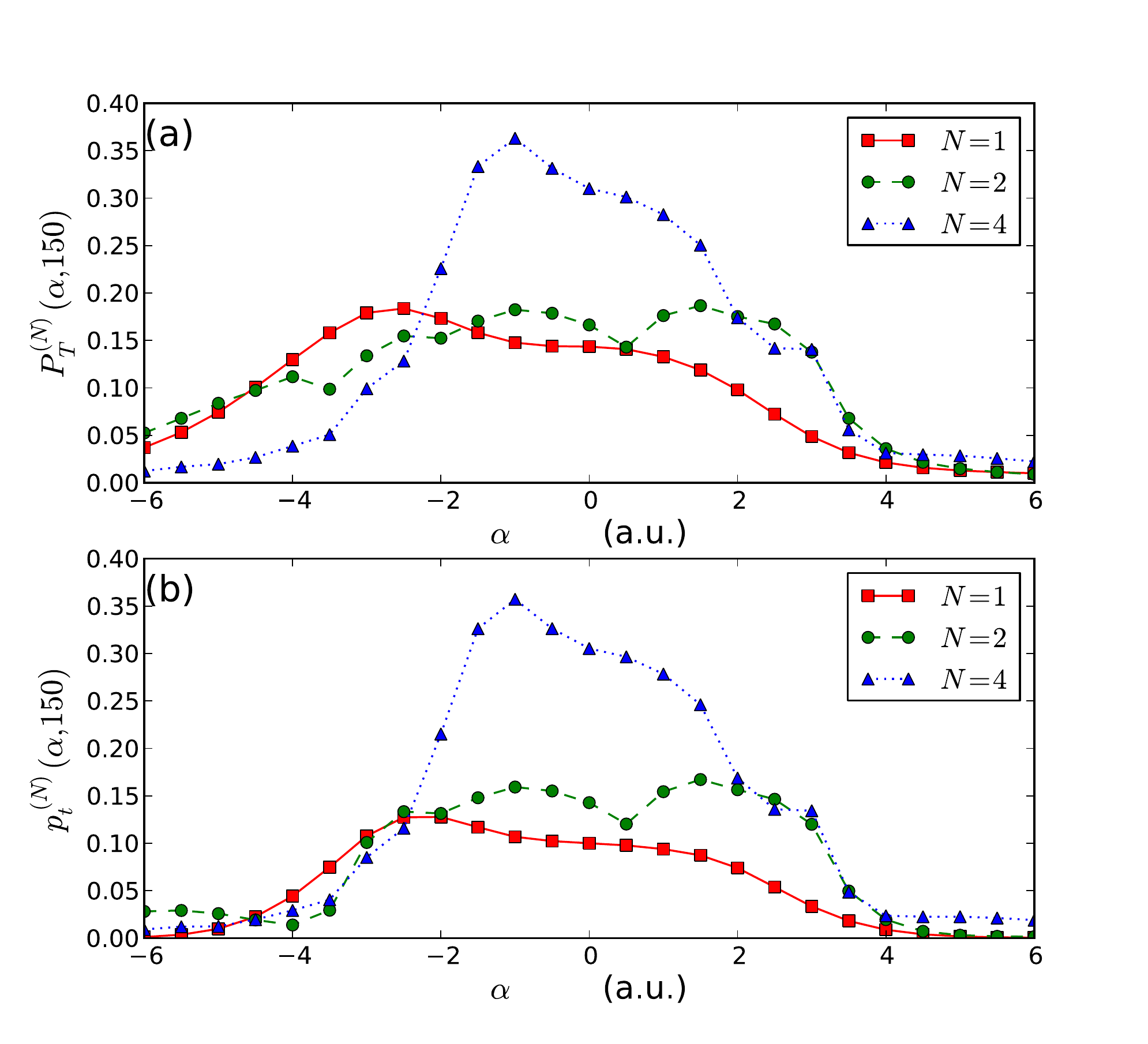}
\caption{(Color online) Probabilities of tunneling as a function of the height of the barrier ($\alpha$). (a) $P_T^{(N)}(\alpha, 150)$ [Eq. (\ref{Prob_T_deff})]; (b) $p_t^{(N)}(\alpha, 150)$ [Eq. (\ref{prob_t_deff})]. }\label{Fig_tunneling_prob}
\end{center}
\end{figure}

\begin{figure}
\begin{center}
\includegraphics[scale=0.45]{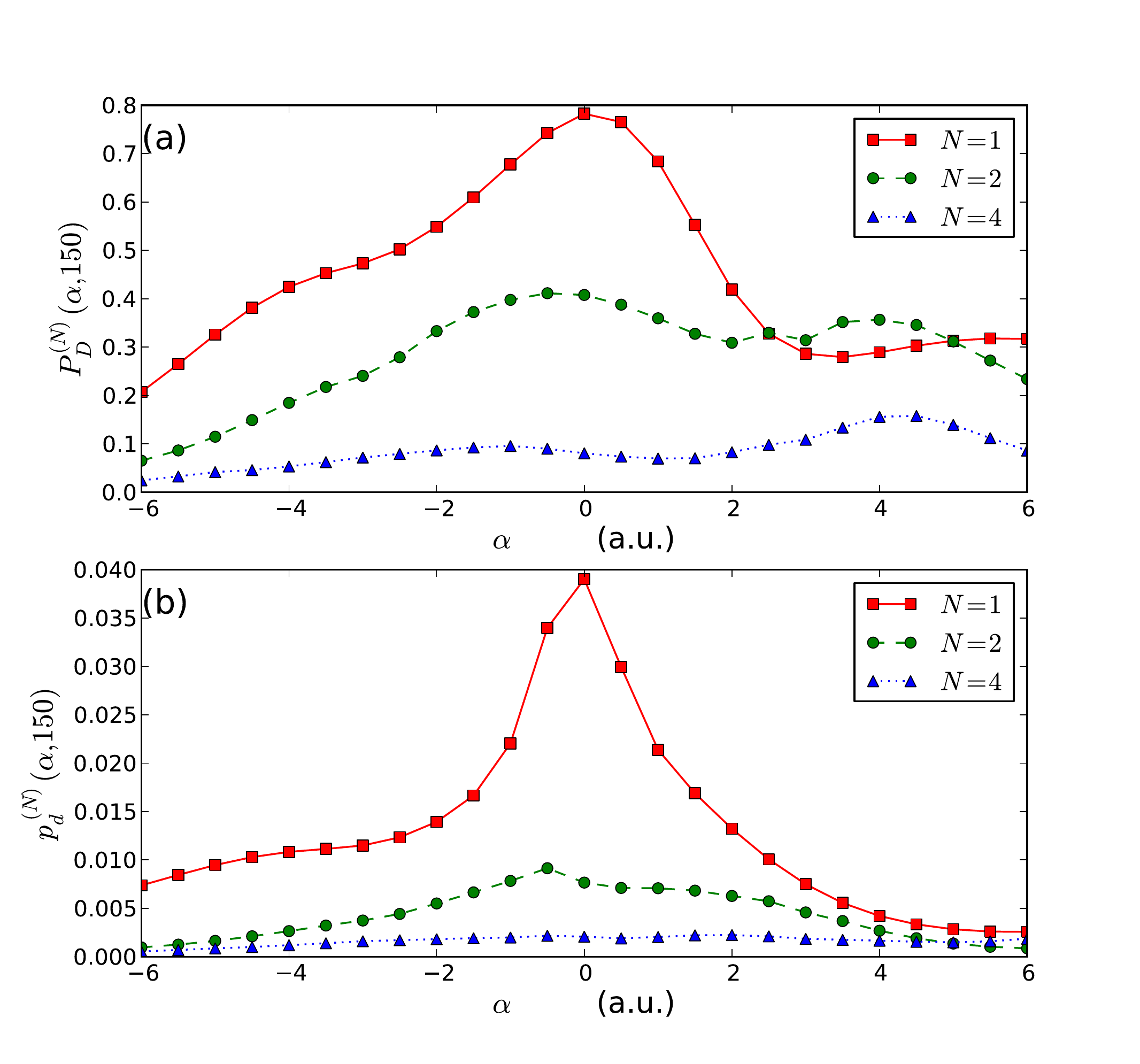}
\caption{(Color online) Probabilities of disintegration as a function of the height of the barrier ($\alpha$). (a) $P_D^{(N)}(\alpha, 150)$ [Eq. (\ref{Prob_D_deff})]; (b) $p_d^{(N)}(\alpha, 150)$ [Eq. (\ref{prob_d_deff})]. }\label{Fig_disintegr_prob}
\end{center}
\end{figure}

\begin{figure}
\begin{center}
\includegraphics[scale=0.45]{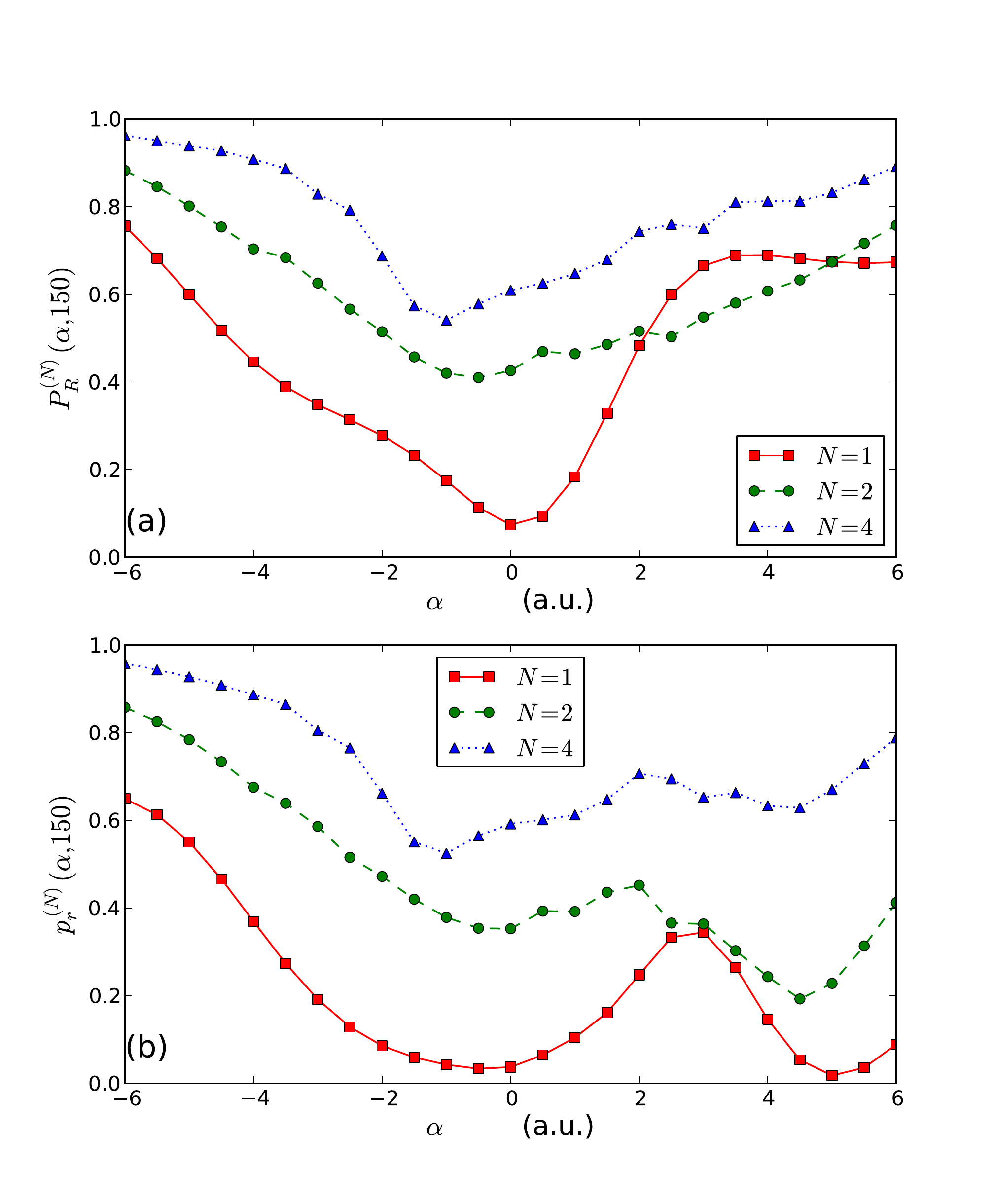}
\caption{(Color online) Probabilities of reflection as a function of the height of the barrier ($\alpha$). (a) $P_R^{(N)}(\alpha, 150)$ [Eq. (\ref{Prob_R_deff})]; (b) $p_r^{(N)}(\alpha, 150)$ [Eq. (\ref{prob_r_deff})]. }\label{Fig_reflection_prob}
\end{center}
\end{figure}

\begin{figure}
\begin{center}
\includegraphics[scale=0.45]{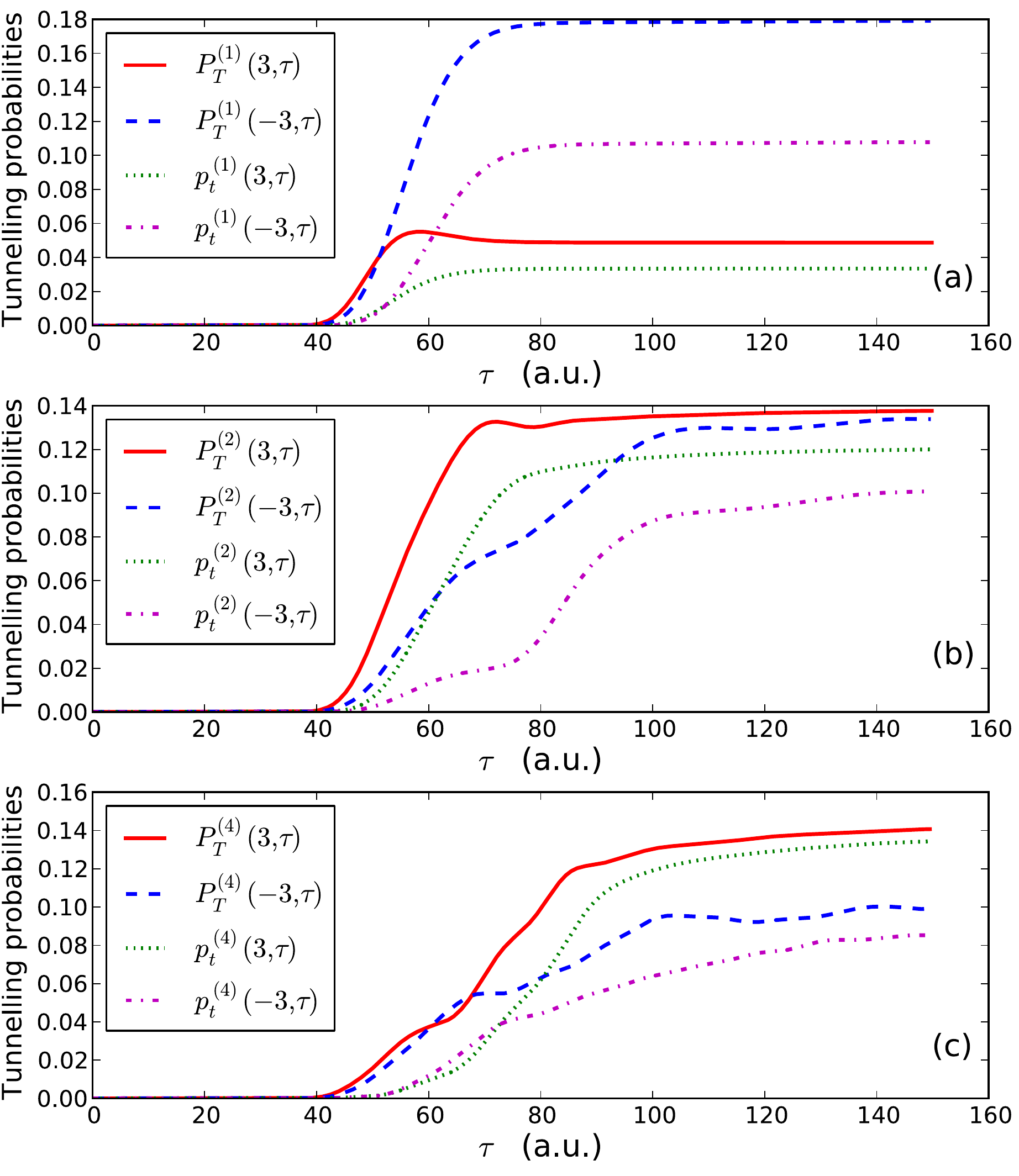}
\caption{(Color online) The probabilities of tunneling $P_T^{(N)} (\alpha, \tau)$ [Eq. (\ref{Prob_T_deff})] and $p_t^{(N)} (\alpha, \tau)$ [Eq. (\ref{prob_t_deff})] as functions of time $\tau$. (a) Comparison of Fig. \ref{Fig_potentials}(a) vs. Fig. \ref{Fig_potentials}(b); (b) comparison of Fig. \ref{Fig_potentials}(c) vs. Fig. \ref{Fig_potentials}(d); (c) comparison of Fig. \ref{Fig_potentials}(e) vs. Fig. \ref{Fig_potentials}(f).}\label{Fig_probability_tunnel_time}
\end{center}
\end{figure}

The results presented in Figs. \ref{Fig_tunneling_prob}--\ref{Fig_probability_tunnel_time} are obtained from the  numerical solution of the  time-dependent Schr\"{o}dinger equation (\ref{TimDepSchEq}) by means of the split-operator method with an absorbing boundary condition. Figures \ref{Fig_tunneling_prob}--\ref{Fig_reflection_prob} show the dependence of the probabilities of tunneling, disintegration, and reflection as function of the parameter $\alpha$ that characterizes the asymmetry of the potential barrier. Dynamics of tunneling processes occurring in Figs. \ref{Fig_potentials}(a)--(f) are visualized as animations, which are available for viewing in Ref. \cite{EPAPS_Animations}. 

Remarkably, while our qualitative conclusion reached regarding Figs. \ref{Fig_potentials}(a) and \ref{Fig_potentials}(b) is indeed correct (see Fig. \ref{Fig_probability_tunnel_time}), the conclusions regarding the comparison of  Figs. \ref{Fig_potentials}(c) and \ref{Fig_potentials}(d) and Figs. \ref{Fig_potentials}(e) and \ref{Fig_potentials}(f) turn out to be completely wrong. In other words, the particle does prefer to ``go'' through the barrier [Figs. \ref{Fig_potentials}(c) and \ref{Fig_potentials}(e)] rather than flying above the barrier [Figs. \ref{Fig_potentials}(d) and \ref{Fig_potentials}(e)]. Furthermore, even though the potentials $\Omega_1 (\pm 3; R, \rho)$ look similar to $\Omega_{2,4} (\pm 3; R, \rho)$,  the particle favors motion above the barrier [Fig. \ref{Fig_potentials}(b)] rather than penetration through the barrier [Fig. \ref{Fig_potentials}(a)] for the former pair of the potentials. This ``paradox'' is resolved in the next section.

\section{Explanation of the effect}\label{Sec3}

The ``paradox'' posed in Sec. \ref{Sec2} is resolved in this section by analyzing a perturbation theory solution of the Schr\"{o}dinger equation. 

In this section, we shall study a general two-1D-particles system given by the Hamiltonian
\begin{eqnarray}\label{HamiltonianR_RHO}
\hat{H} &=& \frac{-1}{2M}\frac{\partial^2}{\partial R^2} + \frac{-1}{2\mu}\frac{\partial^2}{\partial \rho^2} + \nonumber\\
&& U(\rho) + V_1\left( R - \frac{\mu}{m_1} \rho\right) + V_2\left( R + \frac{\mu}{m_2} \rho\right),
\end{eqnarray}
which is already written in the (general) center of mass ($R$) and relative ($\rho$) coordinates,
\begin{eqnarray}
&& R = (m_1 x_1 + m_2 x_2)/(m_1 + m_2), \quad \rho = x_2 -x_1, \nonumber\\
&& \mu = m_1 m_2 /M, \quad M = m_1 + m_2, \nonumber
\end{eqnarray}
where $x_{1,2}$ ($m_{1,2}$), as previously,  being the coordinates (masses) of the first and second particles, respectively.

Let us introduce the following notation 
\begin{eqnarray}
\hat{U}(t_f, t_i) &=& \hat{T} \exp\left[ -i \hat{H}(t_f - t_i)\right], \\
\hat{U}_0(t_f, t_i) &=& \hat{T} \exp\left[ -i( \hat{H} - V_1 - V_2)(t_f - t_i) \right],
\end{eqnarray} 
for total and unperturbed propagators, respectively. The sum of the potential barriers, $V_1 + V_2$, shall be considered as a perturbation. The eigenfunctions, $\ket{n}$, and eigenvalues, $E_n$, of the internal motion are the solutions of the problem 
\begin{eqnarray}\label{InternalMotionEigenstates}
\left[ \frac{-1}{2\mu}\frac{d^2}{d\rho^2} + U(\rho) \right]\ket{n} = E_n\ket{n},
\end{eqnarray}
where the index $n$ denotes bound and continuous states. Introducing $\ket{n k} \equiv \ket{n} \otimes \ket{k}$, where $\ket{k}$ is an eigenfunction of the free motion of the center of mass $\langle R \ket{k} \equiv \exp(ikR)/\sqrt{2\pi}$ -- a plane wave and $\ket{n}$ is an eigenstate of the internal motion, the unperturbed propagator reads
\begin{eqnarray}
\hat{U}_0(t_f, t_i) = \intsum_n \int dk e^{-i\left(E_n + \frac{k^2}{2M}\right)(t_f - t_i)}\ket{n k}\bra{k n}
\end{eqnarray} 
The total propagator is a solution of the Lippmann-Schwinger equation written in the ``post'' form 
\begin{eqnarray}
\hat{U}(t_f, t_i) &=& \hat{U}_0(t_f, t_i) \nonumber\\
&-& i \int_{t_i}^{t_f} dt \hat{U}_0(t_f, t) [V_1 + V_2] e^{\epsilon t} \hat{U}(t, t_i),
\end{eqnarray} 
where we set $\epsilon \to 0$.

Assuming that the initial condition $\ket{\Psi(t_i)} \equiv \ket{n}\otimes\ket{\psi_{in}}$, where $\ket{n}$ is one of the eigenstates (\ref{InternalMotionEigenstates}) and $\ket{\psi_{in}}$ is a wave packet localized before the barriers [see, e.g., Eq. (\ref{InitialCondition})],  we obtain
\begin{eqnarray}
\ket{\Psi(+\infty)} &\approx& \hat{U}_0(+\infty, -\infty)\ket{\Psi(-\infty)} + \ket{\Psi_1} + \ket{\Psi_2}, \label{Psi_at_plus_infty}\\
\ket{\Psi_1} &=& -2\pi i\intsum_{n'} \int dk dk' \delta\left( E_{n'} + \frac{k'^2}{2M} - E_n - \frac{k^2}{2M} \right) \nonumber\\
&& \times \ket{n' k'} W_{n n'}(k-k') \langle k\ket{\psi_{in}}, \\
\ket{\Psi_2} &=& -2\pi i\intsum_{n'', n'} \int dkdk'dk'' \ket{n'' k''}\langle k \ket{\psi_{in}} \nonumber\\
&& \times \delta\left( E_{n''} + k''^2 /[2M] - E_n - k^2/[2M]\right) \nonumber\\
&&\times \frac{ W_{n' n''}(k'-k'') W_{n n'}(k-k')}{E_n + k^2/(2M) - E_{n'} - k'^2/(2M) + i0},
\end{eqnarray}
where $W_{n n'}(k - k') = \bra{k' n'} V_1 + V_2 \ket{n k}$ and $\ket{\Psi_{1,2}}$ are the first- and second-order corrections, respectively. Higher-order corrections can be derived in a similar manner, but what is important for our further analysis is that they are functions of $W$.  

We simplify the matrix element $W$ by representing it as follows
\begin{eqnarray}\label{W_initial_expression}
W_{n n'}(k - k') &=& \int \frac{dR d\rho dq}{2\pi} e^{i(k-k')R}\phi_{n'}^*(\rho)\phi_n(\rho) \nonumber\\
&& \times [ V_1(q)\delta(R - \mu\rho/m_1 - q) \nonumber\\
&& + V_2(q)\delta(R + \mu\rho/m_2 - q)],
\end{eqnarray}
where  $\phi_n(\rho) = \langle \rho \ket{n}$. After trivial integration over $R$, we obtain
\begin{eqnarray}\label{Simplified_W_general}
&& W_{n n'}(k - k') = \mathrsfs{F}_{n n'}\left( \frac{\mu}{m_1}[k-k']\right)\int \frac{dq}{2\pi} e^{i(k-k')q} V_1(q) \nonumber\\
&& \quad + \mathrsfs{F}_{n n'}\left( \frac{\mu}{m_2}[k'-k]\right)\int \frac{dq}{2\pi} e^{i(k-k')q} V_2(q),
\end{eqnarray}
where the quantity, 
\begin{eqnarray}\label{Formfactor_Def}
 \mathrsfs{F}_{n n'}(p) = \int d\rho e^{ip\rho} \phi_{n'}^*(\rho)\phi_n(\rho),
\end{eqnarray}
is called the form factor, and it is well known in the scattering theory. Its physical interpretation is the probability amplitude of transferring a momentum $p$ from the center of mass to the interparticle degree of freedom by making the transition $n\to n'$. 

Now we consider the case of identical particles: $m_1 = m_2 = m$ and $U(-\rho) = U(\rho)$. Then, there are two types of the eigenstates of the internal motion: even ($+$), $\phi_n(-\rho) = \phi_n(\rho)$, and odd ($-$), $\phi_n(-\rho) = -\phi_n(\rho)$. Since $ \mathrsfs{F}_{n n'}(-p) =  \mathrsfs{F}_{n n'}(p)$ [$\mathrsfs{F}_{n n'}(-p) =  -\mathrsfs{F}_{n n'}(p)$] in the case of $\phi_n$ and $\phi_{n'}$ being of the same (different) parity,  
Eq. (\ref{Simplified_W_general}) takes the form
%%%%%%%%%%%%%%%%%%%%%%%%%%%%%%%%%%%%%%%%%%%%%
% Dear Editor, please DO NOT rearrange the following equation into a double page format
% because its central formula of this paper.
%%%%%%%%%%%%%%%%%%%%%%%%%%%%%%%%%%%%%%%%%%%%%
\begin{widetext}
\begin{eqnarray}\label{W_equivalent_particles}
W_{n n'}(k - k')  &=&  \left\{ 
\begin{array}{ccl}
\mathrsfs{F}_{n n'}\left( [k - k']/2\right) \int \frac{dq}{2\pi}e^{i(k-k')q} [V_1(q) + V_2(q)], &\mbox{if}& \mbox{$\phi_n$ and $\phi_{n'}$ have the same parity} \\
\mathrsfs{F}_{n n'}\left( [k - k']/2\right) \int \frac{dq}{2\pi}e^{i(k-k')q} [V_1(q) - V_2(q)], &\mbox{if}& \mbox{$\phi_n$ and $\phi_{n'}$ have different parities}, 
\end{array}
\right.
\end{eqnarray}
\end{widetext}
which is the product of the form factor and the Fourier transform of either the sum of the barriers or the difference of the barriers, depending on the parities of the initial and final states. 

Two conclusions can be readily drawn from Eq. (\ref{W_equivalent_particles}): First, considering tunneling within the time-independent picture, Amirkhanov and Zakhariev \cite{Amirkhanov1966} have discovered the violation of the barrier penetration symmetry for complex particles, i.e., the penetration of composite particles through asymmetric barriers in opposite directions may differ. [Note that the rates of tunneling of an elementary (structureless) particle are exactly the same in both the directions within the time-independent approach.] The situations when the system approaches the barrier from the left and from the right differ only by inversion of the sign of the momentum of the center of mass. The only part of the wave function (\ref{Psi_at_plus_infty}) that maintains the dependence on the sign of the momentum is the matrix element (\ref{Simplified_W_general}).  Thus, the discussed phenomenon of tunneling asymmetry is manifested in our consideration as a physical consequence of the property 
\begin{eqnarray}\label{SymmetryW}
W_{n n'}(k'-k) = W_{n n'}(k-k') \Longleftrightarrow V_1(q) = V_2(-q).
\end{eqnarray}
Equation (\ref{SymmetryW}) is not only an alternative and perhaps faster way of achieving the main result of Ref. \cite{Amirkhanov1966} but also the generalization of their conclusion for the case of nonidentical barriers ($V_1 \neq V_2$).

Second, Eq. (\ref{W_equivalent_particles}) basically provides an explanation of the observed anomalies related to the potentials $\Omega_n (\alpha; x_1, x_2)$ [Eq. (\ref{PotentialDeff})] pictured in Fig. \ref{Fig_potentials}, if we recall that tunneling of a 2D particle in the potential $\Omega_n (\alpha; x_1, x_2)$ is equivalent to collective tunneling of two equal 1D particles through the potential barriers $V_1(x_1) = \alpha V(x_1)$ and $V_2(x_2) = 3 V(x_2)$. Hence, Eq. (\ref{SymmetryW}) determines the selection rule for transitions between states of the internal degree of freedom induced by (collective) tunneling. The key point is that the probability of collective tunneling strongly depends on whether an excitation of the internal degree of freedom is possible. If a system is initially in the ground state and the excitations are allowed, then by going to an excited state, the center of mass of the system lowers its kinetic energy (i.e., increasing the width of the barrier), consequently reducing the probability of tunneling (see Refs. \cite{Zakhariev2002, Goodvin2005, Goodvin2005a, Hnybida2008, Shegelski2008} and reference therein). We recall that the parity of the ground state is even, the first excited state -- odd, the second excited state -- even, etc.; therefore,  according to Eq. (\ref{W_equivalent_particles}), if 
\begin{eqnarray}\label{Condition_not_forcing_transitions}
V_1(q) = V_2(q),
\end{eqnarray} 
then the transition from the ground state to the first excited state is forbidden, but if 
\begin{eqnarray}\label{Condition_forcing_transitions}
V_1(q) = -V_2(q),
\end{eqnarray} 
then the transition is allowed. Paraphrasing, we note that if condition (\ref{Condition_not_forcing_transitions}) takes place then the interparticle degree of freedom may stay the same while the center of mass traverses the barriers, but if condition (\ref{Condition_forcing_transitions}) holds then the state of the interparticle degree of freedom can change. In essence, this is the core of the observed phenomenon in Sec. \ref{Sec2}.

Indeed, condition (\ref{Condition_forcing_transitions}) is satisfied for Figs. \ref{Fig_potentials}(d) and \ref{Fig_potentials}(f). Hence, the tunneling probability is less in these cases than in cases Figs. \ref{Fig_potentials}(c) and  \ref{Fig_potentials}(e) for which equality (\ref{Condition_not_forcing_transitions}) takes place. On the whole, the same conclusion is valid as long as the potential $U_N(\rho)$ can have at lest two bound states. In Figs. \ref{Fig_potentials}(a) and \ref{Fig_potentials}(b), when there is only a single bound state supported by the intraparticle interaction, the simple intuitive picture holds. Why is this the case? After all, there are also continuum states of the intraparticle motion available for the excitation.

To answer this question, let us look at collective tunneling from the point of view of the (time-independent) multichannel  formalism, which is the most common method employed to the problem at hand  (see, e.g., Refs. \cite{Zakhariev1964, Amirkhanov1966, Zakhariev2002, Saito1994, Penkov2000, Chabanov2000, Penkov2000a, Razavy2003a, Goodvin2005, Goodvin2005a, Lee2006, Hnybida2008, Shegelski2008, Zakhariev2008}). According to the multi-channel approach, using the expansion $\Psi(R, \rho) = \intsum_n \phi_n(\rho)\chi_n (R)$, the stationary Schr\"{o}dinger equation, $\hat{H}\Psi(R, \rho) = E\Psi(R, \rho)$,  is reduced to the following system of ordinary differential  equations for unknown functions $\chi_n(R)$:
\begin{eqnarray}\nonumber
\frac{1}{2M} \frac{d^2\chi_n(R)}{dR^2} -  \intsum_{n'} Z_{nn'}(R) \chi_{n'}(R) = (E_n-E)\chi_n(R),
\end{eqnarray}
where $Z_{nn'}(R)$ being effective potentials,
\begin{eqnarray}\label{Effective_Potential_Def}
Z_{nn'}(R) &=& \int d\rho \, \phi^*_n(\rho)\left[ V_1\left( R - \frac{\mu}{m_1} \rho\right) + \right.  \nonumber\\
&& \left. + V_2\left( R + \frac{\mu}{m_2} \rho\right) \right]\phi_{n'}(\rho).
\end{eqnarray}
Such an effective potential may be interpreted as the potential barrier that the center of mass encounters while incident in the state $n'$ and reflected or transmitted in the state $n$. 

\begin{figure}
\begin{center}
\includegraphics[scale=0.45]{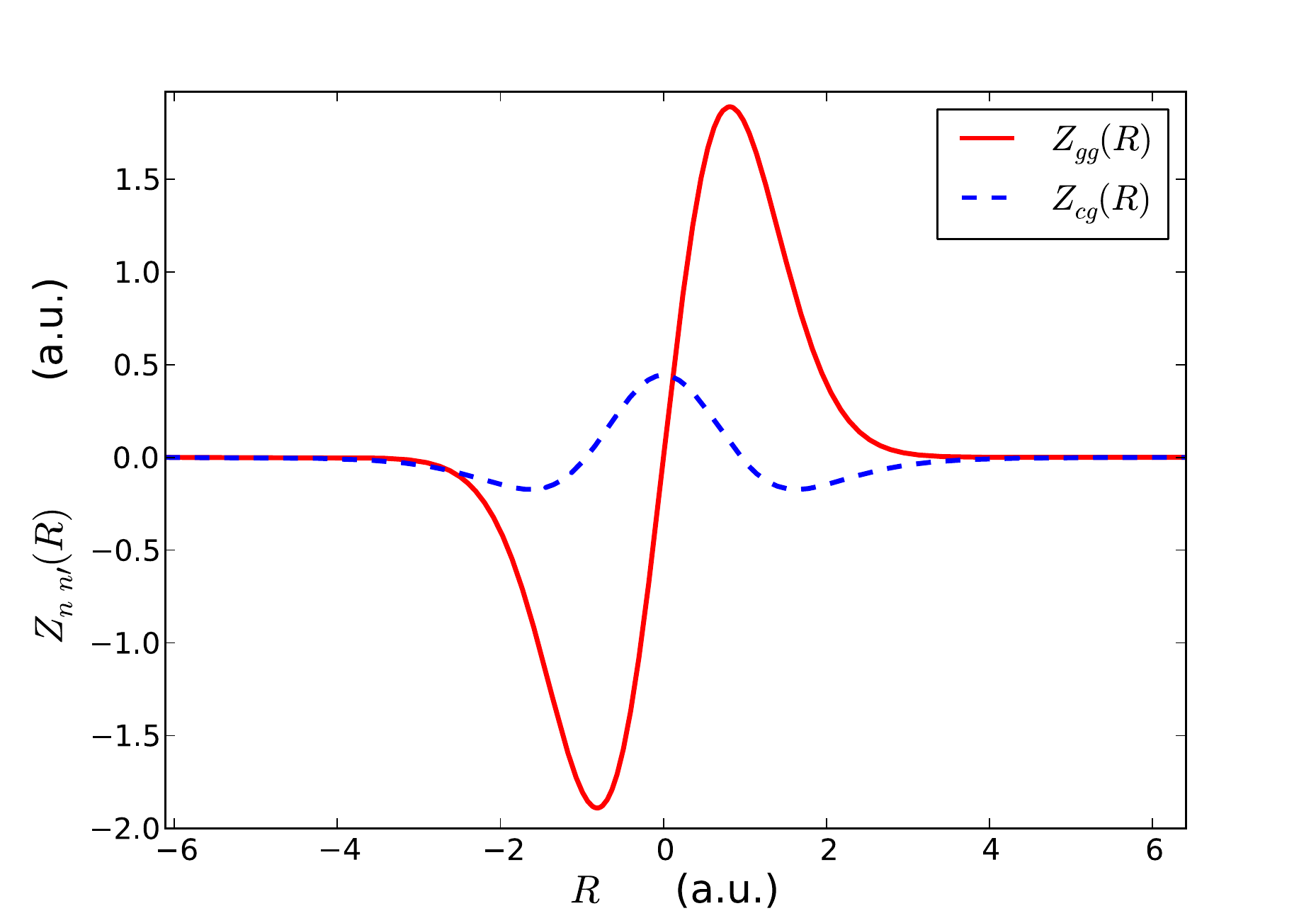}
\caption{(Color online) Plots of effective potentials $Z_{nn'}(R)$ [Eq. (\ref{Effective_Potential_Def})]. The solid line is $Z_{gg}(R)$ for case (a) of Fig. \ref{Fig_potentials}. The dashed line represents $Z_{cg}(R)$ for case (b) of Fig. \ref{Fig_potentials}. }\label{Fig_effective_potentials}
\end{center}
\end{figure}

Let $g$ denote the ground as well as the single  bound state of the potential 
$U_1(\rho)$ [Eq. (\ref{Un_potentials})], and $c$ a low-lying odd (unbound) state of the continuum spectrum, which is normalized to the delta function. Then, $Z_{cg}(R) \equiv Z_{gc}(R) \equiv 0$ since the $g\to c$ transition is forbidden in Fig. \ref{Fig_potentials}(a) ; respectively, $Z_{gg}(R) \equiv 0$ in Fig. \ref{Fig_potentials}(b). The first nonzero effective potentials in Figs. \ref{Fig_potentials}(a) and \ref{Fig_potentials}(b) are $Z_{gg}(R)$ and $Z_{cg}(R)$, respectively, and they are plotted in Fig. \ref{Fig_effective_potentials}. Taking into account that $E_g - \bar{E}_1 \approx E_{cm} = 1$ (a.u.) ($E_g$ being the ground state energy) is the kinetic energy of the center of mass, we may qualitatively conclude from  Fig. \ref{Fig_effective_potentials} that the center of mass needs to tunnel through the barrier  [$Z_{gg}(R)$] in Fig. \ref{Fig_potentials}(a) and flies above the barrier [$Z_{cg}(R)$] in Fig. \ref{Fig_potentials}(b); thus, the probability of finding the particle in the first quadrant in Fig. \ref{Fig_potentials}(b) prevails over the probability of tunneling in Fig. \ref{Fig_potentials}(a). Finally, we note that the first nonzero effective potentials in Figs. \ref{Fig_potentials}(c) and \ref{Fig_potentials}(d) are of the same order; the same statement is valid in cases Figs. \ref{Fig_potentials}(e) and \ref{Fig_potentials}(f).

Traces of the forced excitations, which occur in Figs. \ref{Fig_potentials}(d) and \ref{Fig_potentials}(f), can be directly observed in obtained numerical data; these are steplike structures in $P^{(2)}_T(-3, \tau)$, $p^{(2)}_t (-3, \tau)$, $P^{(4)}_T(-3, \tau)$, and $p^{(4)}_t (-3, \tau)$  [see Figs. \ref{Fig_probability_tunnel_time}(b) and \ref{Fig_probability_tunnel_time}(c)] and   a snakelike shape of the wave function $\Psi_{2,4}(-3; t, R, \rho)$ that emerges from the barrier (see Ref. \cite{EPAPS_Animations}). Nevertheless, the data regarding Fig. \ref{Fig_potentials}(b) [see Fig. \ref{Fig_probability_tunnel_time}(a)] seems not to reveal similar jumps at first sight. It is due to the fact that the coupling between bound states [the form factor (\ref{Formfactor_Def}), more precisely] is bigger than the coupling of a bound state to a state of the continuum spectrum.  Regardless of smallness, these transitions show up in the observation that the probability of disintegration in Fig. \ref{Fig_potentials}(a) is less than in Fig. \ref{Fig_potentials}(b) [see Fig. \ref{Fig_disintegr_prob} that $P_D^{(1)} (3, 150) < P_D^{(1)} (-3, 150)$ as well as $p_d^{(1)} (3, 150) < p_d^{(1)} (-3, 150)$].

Concluding this section, we list pivotal factors in explaining the ``paradox'' reported in Sec.  \ref{Sec2}. (i) This effect cannot exist in one dimension, it requires at least two dimensions. (ii) Our explanation of the effect relies on a natural isomorphism between systems of one 2D particle and two 1D particles of the same mass. The essence of the effect lies in possibility of tuning the potential barriers such that the intraparticle degree of freedom is  excited. (iii) Dynamics of tunneling crucially depends on whether the intraparticle potential supports one or more bound states (an exact number is irrelevant for the qualitative description). 

\section{Classical Physics and The ``paradox''}\label{Sec4}

A classical counterpart of the quantum system at hand [Eq. (\ref{HamiltonianCartesianDef})] is a mechanical system with the Hamiltonian 
\begin{eqnarray}\label{ClassicalHamiltonianCartesian}
\mathrsfs{H}(p_1, p_2; x_1, x_2) = \left( p_1^2 + p_2^2 \right)/2 + \Omega_N (\alpha; x_1, x_2),
\end{eqnarray}
where $p_{1,2}$ and $x_{1,2}$ are canonically conjugate variables. One may perform the canonical transformation to rewrite the Hamiltonian (\ref{ClassicalHamiltonianCartesian}) in terms of the new canonical variables $P_R$, $P_{\rho}$ and $R$, $\rho$, where the latter pair being the center of mass and relative coordinates [Eq. (\ref{SimpleCMandRelativeCoord})],
\begin{eqnarray}\label{ClassicalHamiltonianCM_Relative}
\mathrsfs{H}\left( P_R, P_{\rho}; R, \rho\right) = P_R^2 /4 + P_{\rho}^2 + \tilde{\Omega}_N (\alpha; R, \rho).
\end{eqnarray}
The connection between new and old canonical momenta reads:
$
p_1 = P_R/2 - P_{\rho}, \quad p_2 = P_R/2 + P_{\rho}.
$

The initial condition for the classical counterpart that corresponds to the initial condition (\ref{InitialCondition}) is $x_1(0) = x_2(0) = \bar{R}$ and $p_1(0) = p_1(0) = \left[ \bar{E}_N -\Omega_N (\alpha; \bar{R}, \bar{R}) \right]^{1/2}$. Having calculated classical trajectories with this initial condition, we observe that the classical particle does not reach the first quadrant in Figs. \ref{Fig_potentials}(b), \ref{Fig_potentials}(d), and \ref{Fig_potentials}(f); it is reflected back to the third quadrant. The same conclusion can be reach qualitatively by calculating the force that acts on the classical counterpart, 
\begin{eqnarray}
F_{1,2}(x_1, x_2) = -\partial \Omega_N (\alpha; x_1, x_2) /\partial x_{1,2}.
\end{eqnarray}
Since 
\begin{eqnarray}\nonumber
F_1(x,x) =  (2\alpha x^2 -\alpha) e^{-x^2}, \quad
F_2(x,x) = (6x^2 - 3)e^{-x^2},
\end{eqnarray}
the classical particle experiences the force that deflects it from moving along the diagonal, $x_1 = x_2$ (which coincides with the axis $\rho=0$), and pushes it toward a knee-like barrier located in the second quadrant (see Fig. \ref{Fig_potentials}); hence, the particle eventually bounces off the barrier back to the third quadrant. 

It is noteworthy to mention a peculiarity of numerical calculations. We have found that it is advantageous to employ a (fourth-order) symplectic integrator \cite{Forest1990, Yoshida1990, Candy1991} for solving Hamilton's equations in this section due to the following reason: A sharp and localized shape of the kneelike barrier leads to an unstable motion of the classical particle. If one employs nonsymplectic integrators (e.g., the Runge-Kutta methods), a very tiny time step must be chosen in order to properly account for the influence of the kneelike barrier; this, in fact, often leads to instability of the numerical scheme for a long time propagation. A physical reason of such an instability lies in the fact that nonsymplectic integrators do not explicitly conserve energy while the symplectic integrators always do; hence, they give a proper long-time evolution of any chaotic Hamiltonian system.

The observation of this behavior of the classical counterpart casts doubt on the quantum nature of the ``paradox.'' More precisely, is it possible that an ensemble of classical particles, which corresponds (in some sense) to the initial wave function of the system at hand (\ref{InitialCondition}), would mimic the observed phenomenon? As shown below, the answer turns out to be negative. 

Furthermore, since it is well known that the application of the semi classical approximation, as a mediator between classical and quantum mechanics, to tunneling often is very fruitful in shading light on the physical nature of the studied process (see, e.g., Refs. \cite{Maitra1997, Spanner2003}), the addressed question is important as the first step toward the usage of semiclassical methods for the interpretation of the ``paradox.''

Shirokov has proposed the unified formalism for quantum and classical mechanics \cite{Shirokov1979d}--a reformulation of both the theories in terms of the same physical and mathematical concepts. Crudely speaking, 
this formalism is based on the well-known fact that observables of quantum mechanics can be converted from operators to functions (i.e., to a very similar form as in classical physics) by means of the Weyl representation \cite{Weyl1950}. In these terms, the probability density of quantum states is represented by the Wigner quasiprobability density distribution function \cite{Wigner1932}. Hence, in order to construct a classical ensemble that corresponds to the quantum particle, we shall calculate the Wigner function $W(P_R, P_{\rho}; R, \rho)$ for the initial condition (\ref{InitialCondition}), 
\begin{eqnarray}\label{WingerFunctionForInitialState}
&& W(P_R, P_{\rho}; R, \rho) = (2\pi)^2 \int \Psi\left( R-R'/2, \rho-\rho'/2\right)\\
&& \qquad\quad \times \Psi^*\left( R+R'/2, \rho+\rho'/2\right) e^{i\left( P_R R'\ + P_{\rho} \rho' \right)} d\rho' dR', \nonumber
\end{eqnarray}
where $\Psi (R, \rho) \equiv \Psi_N(\alpha; 0, R, \rho)$. Since $\phi_g(\rho)$ has no zeros and decays exponentially at infinity, we shall approximate it by a Gaussian 
\begin{eqnarray}\label{GaussApproxGroundState}
\phi_g(\rho) \propto \exp\left[ -\rho^2 / \left(2\sigma_{\rho}^2\right) \right],
\end{eqnarray}
where we set $\sigma_{\rho} = 1.5$ (a.u.). Substituting Eqs. (\ref{GaussApproxGroundState}) and (\ref{InitialCondition}) into Eq. (\ref{WingerFunctionForInitialState}), one readily obtains
\begin{eqnarray}
W(P_R, P_{\rho}; R, \rho)  &\propto& \exp\left\{ -(R-\bar{R})^2 / \sigma_R^2 -\rho^2 / \sigma_{\rho}^2 \right. \\
&& \left. - \sigma_R^2 \left( P_R - \sqrt{2M E_{cm}}\right)^2 -\sigma_{\rho}^2 P_{\rho}^2 \right\}. \nonumber
\end{eqnarray}
Therefore, our quantum system corresponds to an ensemble of classical particles with the Hamiltonian (\ref{ClassicalHamiltonianCM_Relative}), and the initial state of the ensemble that corresponds to the initial condition (\ref{InitialCondition}) can be generated by considering $R$, $\rho$, $P_R$, and $P_{\rho}$ as independent normal random variables with means $\bar{R}$, 0, $\sqrt{2M E_{cm}}$, 0 and with standard deviations $\sigma_R/\sqrt{2}$, $\sigma_{\rho}/\sqrt{2}$, $\left[\sqrt{2}\sigma_R\right]^{-1}$, $\left[\sqrt{2}\sigma_{\rho}\right]^{-1}$, respectively.

\begin{figure*}
\begin{center}
\includegraphics[scale=0.65]{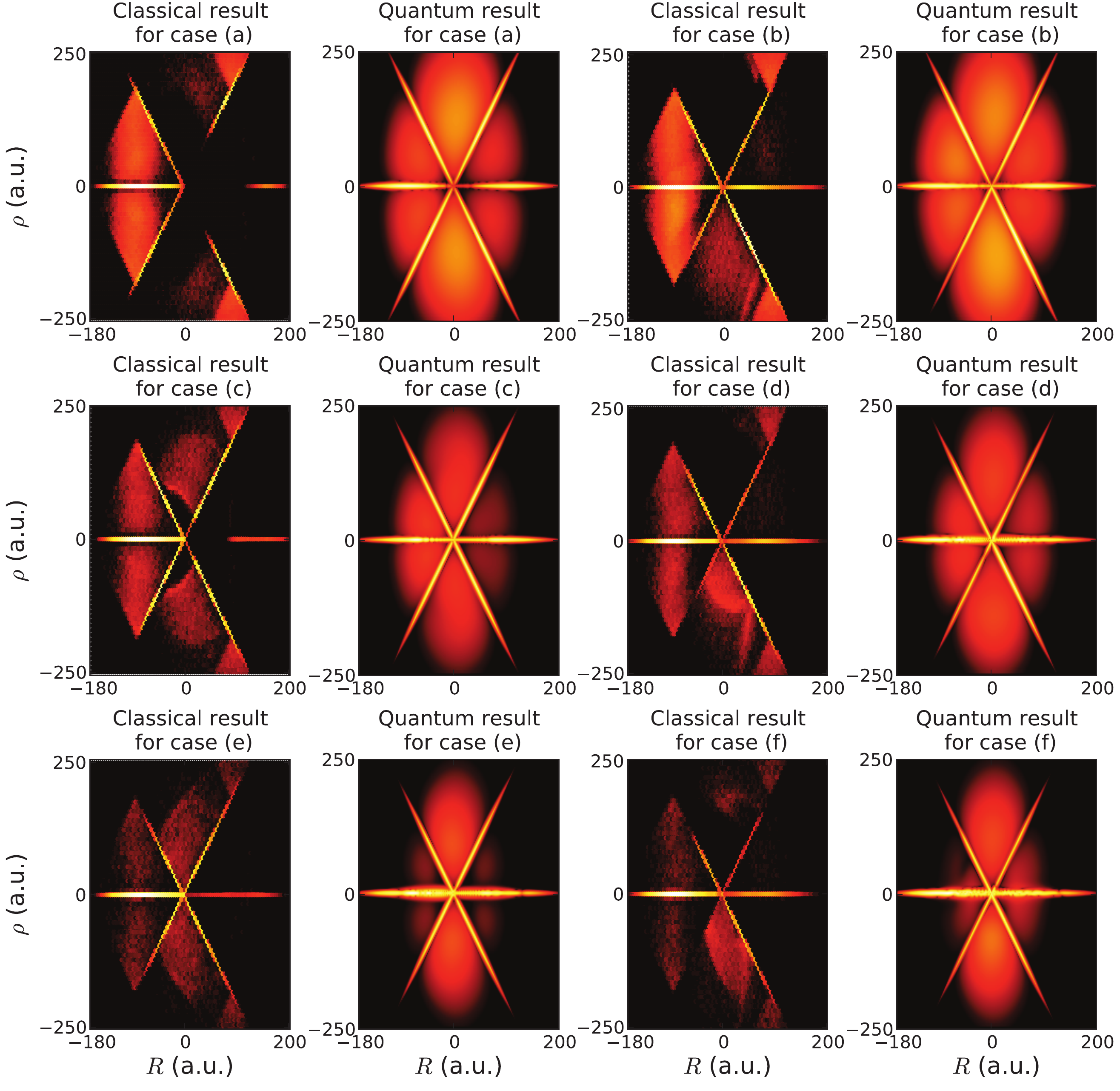}
\caption{(Color online) Classical vs. quantum mechanics for potentials plotted in Fig. \ref{Fig_potentials}. Classical and quantum normalized probability density distributions (in the logarithmic scale) at the time moment $t=150$ (a.u.). Black corresponds to the minimum values, whereas white corresponds to the maximum.}\label{CvsQ}
\end{center}
\end{figure*}

Results of classical simulations of dynamics of an ensemble of $10^6$ particles are compared with {\it ab initio} quantum simulations in Fig. \ref{CvsQ}. Foremost, one may notice how well the classical simulations reproduced quantum behavior in the classically allowed regions in almost all the cases. One observes a qualitative agreement between classical and quantum results in the cases of tunneling [Figs. \ref{CvsQ}(a), \ref{CvsQ}(c), and \ref{CvsQ}(e)]. Nevertheless, in the cases of the over barrier motion [Figs. \ref{CvsQ}(b), \ref{CvsQ}(d), and \ref{CvsQ}(f)], classical mechanics gives more asymmetric probability distributions than quantum mechanics. This can be explained by means of a simple observation that there are the kneelike potential barriers in Figs. \ref{Fig_potentials}(b), \ref{Fig_potentials}(d), and \ref{Fig_potentials}(f), which force a majority of classical particles to go to the fourth quadrant.

Having calculated the classical probability density distributions, we may introduce the probability of tunneling as well as the shifted probability of tunneling analogously to the corresponding quantum quantities [Eqs. (\ref{Prob_T_deff}) and (\ref{prob_t_deff})]. The classical probabilities of tunneling in the above-barrier cases are an order of magnitude larger than the corresponding classical probabilities in the under-barrier cases. This conclusion contradicts the results of the quantum calculations (Fig. \ref{Fig_probability_tunnel_time}). In other words, there is no ``paradox'' in classical physics. It is natural since the ensemble of classical particles should ``prefer'' going above than ``tunneling through'' the barrier. Hence, we have confirmed that the reported effect is genuinely quantum mechanical.     

\section{Conclusions and discussions}

In Sec. \ref{Sec2}, we presented the 2D systems, whose potentials are plotted in Fig. \ref{Fig_potentials}, which hold the unexpected property that the probability of tunneling through a barrier is larger than the probability of flying above a barrier. As it was clarified in Sec. \ref{Sec3}, this phenomenon occurs due to a specific symmetry of the potential [Eq. (\ref{Condition_forcing_transitions})] that forces excitations of an interparticle degree of freedom, thus lowering the probability of tunneling. This effect is overlooked by the intuitive conclusion which uses the language of trajectories within the quasi classical approximation  that the tunneling is an ``exponentially harder'' process than flying above a barrier.  First and foremost, we note that the quasi classical approximation, being an elegant and insightful approach in 1D, is in fact very cumbersome and quite often impractical in 2D. Hence, in most situations of interest different modifications of the original quasi classical approximation that make additional assumptions on the wave function are employed (see, e.g., Ref. \cite{Razavy2003a} and references therein). From this point of view, we conclude that a quasiclassical model capable of explaining the reported ``paradox'' must not only rely on the language of trajectories but also include the quantum transitions that are at the core of the effect.

An important undiscussed issue is the dependence of the reported effect on the initial condition (\ref{InitialCondition}).  If we substitute $\phi_g(\rho)$ in Eq. (\ref{InitialCondition}) by the wave function of the first excited state of the interparticle Hamiltonian in the cases of $N=2$ and $N=4$ (note that $E_{cm}$ must be appropriately decreased such that it would be possible to talk about tunneling), then one may expect that the ``paradox'' should disappear, and one would observe a conventional situation: the probability of tunneling through the barrier [Figs. \ref{Fig_potentials}(c) and \ref{Fig_potentials}(e)] would be smaller than the probability of flying above the barrier [Figs. \ref{Fig_potentials}(d) and \ref{Fig_potentials}(f)]. Indeed, since the transition from the first excited state to the ground state is allowed because condition (\ref{Condition_forcing_transitions}) is satisfied in Figs. \ref{Fig_potentials}(d) and \ref{Fig_potentials}(f), then after making such a jump, the center of mass gains the energy difference; hence, it can more easily tunnel in Figs. \ref{Fig_potentials}(d) and \ref{Fig_potentials}(f) than in Figs. \ref{Fig_potentials}(c) and \ref{Fig_potentials}(e) where this transition is forbidden. 

As far as applications of the effect to quantum control are considered, consider a system of two neutral atoms that interact through the dipole-dipole interaction and are trapped, e.g., by a dipole trap. The magnitudes of the atomic dipoles depend on internal states occupied by the atoms. The internal states of the atoms can be changed for each atom independently by means of a laser with an appropriately tuned frequency, assuming that the atoms have different spectra. Performing such  excitations, we may be able to switch between the cases where either condition (\ref{Condition_not_forcing_transitions}) or condition (\ref{Condition_forcing_transitions}) is valid. Hence, we may allow or forbid the two atomic system to tunnel through the trap.

A generalization of Eq. (\ref{W_equivalent_particles}) as well as Eqs. (\ref{SymmetryW}), (\ref{Condition_not_forcing_transitions}), and (\ref{Condition_forcing_transitions}) to the case of $n$ ($n\geqslant 3$) particles is a nontrivial question that should be addressed in the future. One might expect that such a generalization of  the effect may reveal many new varieties of the phenomenon, which could be interesting from the point of view of  quantum control of tunneling of complex systems. 

\acknowledgments

The authors thank Michael Spanner for fruitful comments. 
D.I.B. acknowledges the Ontario Graduate Scholarship program for financial support. 
M.Yu.I. and W.K.L. acknowledge support of NSERC discovery grants.

\bibliography{2D_Tunnelling}
\end{document}